\documentclass[reprint,amsmath,amssymb,aps]{revtex4-2}
\usepackage[utf8]{inputenc}
\usepackage{dcolumn}
\usepackage{graphicx,dcolumn,bm,braket} \usepackage{xcolor} \usepackage{soul}
\usepackage[colorlinks]{hyperref}
\hypersetup{colorlinks=true,citecolor=blue,linkcolor=blue,urlcolor=blue}
\begin{document}
    \title{Optical representation of thermal nuclear fluctuation effect on band-gap renormalization}
    \author{Kohei Ishii}
    \email{kishii@issp.u-tokyo.ac.jp}
    \author{Jun Haruyama}
    \author{Osamu Sugino}
    \affiliation{Institute for Solid State Physics, The University of Tokyo, Kashiwa, Chiba 277-8581, Japan}
    \date{\today}
    \begin{abstract}
        The bandgap of insulating materials is renormalized in various ways by the electron-phonon interaction owing to the dynamical and quantum fluctuations of nuclei. These fluctuation effects are considered in the perturbative Allen-Heine-Cardona theory using the formulae for the Fan-Migdal and Debye-Waller terms. However, the material dependence is not clear in the formulae. Thus, in this study, we focus on the analytical form of the Debye-Waller term and find that the term can be reformulated using the momentum matrix. In addition, the optical selection rule is found to play a role. For diamond-type materials, the Debye-Waller term can be approximately decomposed into a product of the optical transition energy, the mean square displacement of nuclei, and the dipole transition probability. The decomposition can also be applied with an additional approximation to zinc-blende-type materials, as revealed by our first-principles calculation. The magnitudes of the Debye-Waller term of several materials can thus be estimated using basic physical quantities prior to performing the calculation of the electron-phonon interaction.
    \end{abstract}
    \maketitle
    \section{\label{sec:1}Introduction}
    The electron-phonon interaction (EPI) plays an important role in many areas of condensed matter physics such as semiconductor physics and superconductivity \cite{2017Giustino}. Considerable EPI effects can be observed in the bandgaps of solids at zero temperature and in the temperature dependence of these bandgaps. For example, the effect, or the bandgap renormalization, on diamond amounts to a few hundred meV, as measured spectroscopically in the temperature range of 100--700 K. Further, it is 320--450 meV even at zero temperature due to the zero-point motion of nuclei~\cite{1992Logothetidis}. Therefore, the accurate prediction of the bandgap requires a precise calculation of the EPI in addition to the calculation of the optical transition energy without including the EPI. Consequently, first-principles prediction had started approximately a decade after the 1990s when the density functional perturbation theory of the EPI was established \cite{2001Baroni} even though basic electronic structure theory was already established in the 1980s. Thus, the electron-phonon renormalization (EPR) is a relatively new topic in the field of first-principles calculation.

    Prior first-principles calculations have been categorized as perturbative, which are based on the Allen-Heine-Cardona (AHC) theory developed around 1980 \cite{1976Allen,1981Allen,1983Allen}, and non-perturbative, where nuclear configurations are displaced stochastically with a Monte Carlo sampling technique \cite{2013Patrick,2015Zacharias}, determined as a single thermally distorted configuration~\cite{2016Zacharias, 2020Zacharias}, or described with a path integral molecular dynamics simulation \cite{2004DellaSala}. The AHC theory can be traced back to 1951 when Fan studied the temperature dependent EPR by including lattice vibrations perturbatively \cite{1951Fan} and to 1955 when Anton\v{c}\'{i}k proposed to alternatively include the degree of thermal nuclear vibration as an effective parameter called the Debye-Waller factor \cite{1955Antoncik}. These two methods were unified by Allen, Heine, and Cardona between 1976 and 1983 \cite{1976Allen,1981Allen,1983Allen} within the rigid-ion approximation (RIA), wherein potential sensed by an electron is decomposed into the contributions from each nucleus. The Fan's and Anton\v{c}\'{i}k's terms can be derived as the same perturbation order terms, and they are presently called the Fan-Migdal (FM) and Debye-Waller (DW) terms, respectively \cite{2017Giustino}.
    Further, the AHC theory was first applied to diamond in 2010 \cite{2010Giustino} and subsequently to various semiconductors, such as GaN, Si, BN, AlN, P, LiF, MgO, LiNbO$_3$, AlAs, AlP, AlSb, GaP, and SiC \cite{2014Kawai,2015Ponce,2015Friedrich,2015Antonius,2016Villegas,2020Engel}.
    
    The off-diagonal terms of the electronic self-energy were neglected in these studies but were recently made accessible for the first-principles calculation by the work of Lihm and Park \cite{2020Lihm}. They enabled this by rewriting the DW term with the momentum operator considering the translational invariance of the one-electron potential, along with enabling the consideration of the hybridization of electronic states. Consequently, this method was implemented in open-source software for first-principles calculations: Quantum Espresso \cite{2009Giannozzi,2017Giannozzi}.

    In this paper, we further rewrite the formulation of Lihm and Park \cite{2020Lihm} to transparently relate the DW term with optical transitions. This is realized by decomposing the DW term into terms contributed by each orbital and subsequently rewriting the orbital-resolved Debye-Waller (ORDW) terms using the momentum matrix. Consequently, the optical selection rule (OSR) plays a role in characterizing the DW term, which allows us to describe the DW term of diamond-type materials as a sum of the products of the optical transition energy, the mean square displacement (MSD) of nuclei, and the dipole transition probability under a plausible approximation of the lattice vibrations. Moreover, the approximate decomposition can be applied to zinc-blende-type materials, as revealed by our first-principles calculation. Thus, it is possible to estimate the magnitude of the DW term using basic physical quantities prior to performing the calculation of the EPI.

    The structure of this paper is as follows: In Sec. \ref{sec:2}, the AHC theory (\ref{subsec:2A}), the ORDW term and OSR (\ref{subsec:2B}), and computational details (\ref{subsec:2C}) are explained. In Sec. \ref{sec:3}, we present an approximate form of the DW term (\ref{subsec:3A}), the numerical validation of the bandgap EPR, and the origin of the material dependence of the ORDW term (\ref{subsec:3B}). In Sec. \ref{sec:4}, we present the conclusion of this paper.

    \section{\label{sec:2}The EPR of the electronic structures}
    In this paper, we primarily follow the notation of Ref.~\cite{2017Giustino} for the Born-von-K\'{a}rm\'{a}n boundary condition, Born-Oppenheimer approximation, Kohn-Sham (KS) method, and the harmonic approximation to lattice vibrations, which are briefly described as follows: We denote the $\alpha$-th component of the position of a supercell as $T_\alpha$ or, in the vector notation, as $\mathbf{T}$. The position of a $p$-th unit cell ($p=1,\cdots,N_p$) is denoted as $\mathbf{R}_p$ and the equilibrium position of the $\kappa$-th nucleus in the unit cell is denoted as $\boldsymbol{\tau}_{\kappa p}^0$, which can be decomposed as $\boldsymbol{\tau}_{\kappa p}^0=\mathbf{R}_p+\boldsymbol{\tau}_\kappa^0$. The displacement from $\boldsymbol{\tau}_{\kappa p}^0$ is denoted as $\Delta{\hat{\boldsymbol{\tau}}}_{\kappa p}$ when it is treated as the position operator. Further, we denote the $n$-th KS state as $\left.|\psi_{n\mathbf{k}}\right\rangle$, where $\mathbf{k}$ is the Bloch wave-vector in the first Brillouin zone (FBZ), and the periodic part of $\left.|\psi_{n\mathbf{k}}\right\rangle$ as $\left.|u_{n\mathbf{k}}\right\rangle$. The KS equation is expressed as
    \begin{align} \label{eq:1}
        \left[-{\hat{\nabla}}^2+V^\mathrm{KS}\left(\hat{\mathbf{r}}; \left\{\boldsymbol{\tau}_{\kappa p}^0\right\}\right)\right]\left.|\psi_{n\mathbf{k}}\right\rangle=\varepsilon_{n\mathbf{k}}\left.|\psi_{n\mathbf{k}}\right\rangle,
    \end{align}
    where $\varepsilon_{n\mathbf{k}}$ is the KS eigenvalue and $V^\mathrm{KS}\left(\hat{\mathbf{r}};\left\{\boldsymbol{\tau}_{\kappa p}^0\right\}\right)$ is the effective KS potential for the equilibrium nuclear configuration:
    \begin{align} \label{eq:2}
    	\begin{aligned}
    		V^\mathrm{KS}\left(\mathbf{r};\left\{\boldsymbol{\tau}_{\kappa p}^0\right\}\right)=&\sum_{\kappa,p,\mathbf{T}}{V_\kappa\left(\mathbf{r}-\boldsymbol{\tau}_{\kappa p}^0-\mathbf{T}\right)} \\
    		&+\sum_{\mathbf{T}}\int_\mathrm{sc}\frac{d\mathbf{r}^\prime n\left(\mathbf{r}^\prime;\left\{\boldsymbol{\tau}_{\kappa p}^0\right\}\right)}{\left|\mathbf{r}-\mathbf{r}^\prime-\mathbf{T}\right|} \\
    		&+\left.\frac{\delta E^{xc}\left[n\right]}{\delta n}\right|_{n\left(\mathbf{r};\left\{\boldsymbol{\tau}_{\kappa p}^0\right\}\right)},
    	\end{aligned}
    \end{align}
    where $V_\kappa$ is the pseudopotential of the nucleus $\kappa$, sc means ``supercell'', $n\left(\mathbf{r};\left\{\boldsymbol{\tau}_{\kappa p}^0\right\}\right)$ is the electron density at the equilibrium configuration, and $E^{xc}\left[n\right]$ is the exchange-correlation energy of $n$.
    $V_\kappa$ is a nonlocal pseudopotential, but nonlocality is not important in this paper, so it is written as a function of $\mathbf{r}$ for simplicity of notation.
    The inter-atomic force constant is denoted as
    \begin{align} \label{eq:3}
        C_{\kappa\alpha p,\kappa^\prime\alpha^\prime p^\prime}=\left.\frac{\partial^2E_\mathrm{tot}\left(\left\{\boldsymbol{\tau}_{\kappa p}\right\}\right)}{\partial\tau_{\kappa\alpha p}\partial\tau_{\kappa^\prime\alpha^\prime p^\prime}}\right|_{\{\boldsymbol{\tau}_{\kappa p}^0\}},
    \end{align}
    where $E_\mathrm{tot}$ is the total-energy, and the dynamical matrix is denoted as
    \begin{align} \label{eq:4}
        D_{\kappa\alpha,\kappa^\prime\alpha^\prime}\left(\mathbf{q}\right)=\left(M_\kappa M_{\kappa^\prime}\right)^{-1/2}\sum_{p} C_{\kappa\alpha0,\kappa^\prime\alpha^\prime p}\exp{\left(i\mathbf{q}\cdot\mathbf{R}_p\right)},
    \end{align}
    where $M_\kappa$ is the mass of nucleus $\kappa$. Then, the eigenvalue equation of a phonon is written as
    \begin{align} \label{eq:5}
        \sum_{\kappa^\prime,\alpha^\prime}{D_{\kappa\alpha,\kappa^\prime \alpha^\prime}\left(\mathbf{q}\right)e_{\kappa^\prime\alpha^\prime,\nu}\left(\mathbf{q}\right)}=\omega_{\mathbf{q}\nu}^2e_{\kappa\alpha,\nu}\left(\mathbf{q}\right),
    \end{align}
    where $e_{\kappa\alpha,\nu}\left(\mathbf{q}\right)$ and $\omega_{\mathbf{q}\nu}$ are the eigenvector and eigenvalue, respectively, of the phonon at a wave-vector $\mathbf{q}$ in FBZ and a branch $\nu$.

    \subsection{\label{subsec:2A}The AHC theory}
    In the AHC theory, the EPR of the one-electron energy $\varepsilon_{n\mathbf{k}}$ at temperature $T$ is described as a sum of the FM term and DW term by
    \begin{align} \label{eq:6}
        \Delta\varepsilon_{n\mathbf{k}}\left(T\right)=\Delta\varepsilon_{n\mathbf{k}}^\mathrm{FM}\left(T\right)+\Delta\varepsilon_{n\mathbf{k}}^\mathrm{DW}\left(T\right),
    \end{align}
    as derived by neglecting the off-diagonal terms of the electron self-energy.
    The FM term corresponds to the dynamical correction to $\varepsilon_{n\mathbf{k}}$ and has a form
    \begin{align} \label{eq:7} \begin{aligned}
            \Delta \varepsilon_{n \mathbf{k}}^{\mathrm{FM}}(T)=& \mathrm{Re} \frac{1}{N_p}
            \sum_{m}\sum_{\mathbf{q},\nu}\sum_{\kappa,\alpha,\kappa^\prime,\alpha^\prime}
            l_{\mathbf{q}\nu
            }^2\frac{M_p}{\sqrt{M_\kappa M_{\kappa'}}} \\
        	&\times
            e_{\kappa \alpha, \nu}^*(\mathbf{q}) e_{\kappa^{\prime} \alpha^{\prime}, \nu}(\mathbf{q})
            \left[h_{m n, \kappa \alpha}(\mathbf{k}, \mathbf{q})\right]^{*} \\
            &\times h_{m n, \kappa^{\prime} \alpha^{\prime}}(\mathbf{k}, \mathbf{q})
            \left[ \frac{1-f_{m\mathbf{k+q}}(T)+n_{\mathbf{q} \nu}(T)}{\varepsilon_{n\mathbf{k}}-\varepsilon_{m \mathbf{k}+\mathbf{q}}-\omega_{\mathbf{q}\nu}+i \eta} \right. \\
            &\left. + \frac{f_{m\mathbf{k+q}}(T)+n_{\mathbf{q} \nu}(T)}{\varepsilon_{n\mathbf{k}}-\varepsilon_{m \mathbf{k}+\mathbf{q}}+\omega_{\mathbf{q}\nu}+i \eta}
            \right].
    \end{aligned}\end{align}
    Here, $l_{\mathbf{q}\nu}=(2M_p\omega_{\mathbf{q}\nu})^{-1/2}$ is the zero-point length, $M_p$ is the reference mass that is considered as the proton mass, and $\eta$ is the convergence factor. $f_{m\mathbf{k}+\mathbf{q}}\left(T\right)=\left[e^{(\varepsilon_{m\mathbf{k}+\mathbf{q}}-\varepsilon_F)/k_BT}+1\right]^{-1}$ and $n_{\mathbf{q}\nu}\left(T\right)=\left(e^{\omega_{\mathbf{q}\nu}/k_BT}-1\right)^{-1}$ are the occupation numbers of electrons and phonons, respectively.
  	 $h_{mn,\kappa\alpha}(\mathbf{k},\mathbf{q})$ is defined as
	\begin{align} \label{eq:8}
        h_{mn,\kappa\alpha}\left(\mathbf{k},\mathbf{q}\right)&={\left\langle u_{m\mathbf{k}+\mathbf{q}}\middle|\partial_{\kappa\alpha,\mathbf{q}}{\hat{V}}^\mathrm{KS}\middle| u_{n\mathbf{k}}\right\rangle}_\mathrm{uc},
	\end{align}
	where $\partial_{\kappa \alpha, \mathbf{q}} \hat{V}^\mathrm{KS}$ is defined by
	\begin{align} \label{eq:9}
        \partial_{\kappa\alpha,\mathbf{q}}V^\mathrm{KS}\left(\mathbf{r}\right)&=\sum_{p} e^{-i\mathbf{q}\cdot\left(\mathbf{r}-\mathbf{R}_p\right)}\left.\frac{\partial V^\mathrm{KS}\left(\mathbf{r};\left\{\boldsymbol{\tau}_{\kappa p}^0\right\}\right)}{\partial\tau_{\kappa\alpha}^0}\right|_{\mathbf{r}-\mathbf{R}_p},
	\end{align}
	and uc indicates that the integration is performed over the unit cell.
    Further, the DW term originates from the static correction to the one-electron potential, which arises from the distribution of nuclei around their equilibrium positions. According to Ref.~\cite{2020Lihm}, it has the following form under the RIA~\cite{2014Ponce}:
    \begin{align} \label{eq:10}
    	\begin{aligned}
    		\Delta \varepsilon_{n\mathbf{k}}^\mathrm{DW}(T)=& \frac{i}{N_p}\sum_{\mathbf{q},\nu} \sum_{\kappa,\alpha,\alpha'}
    		\bra{u_{n\mathbf{k}}} \left[ \partial_{\kappa \alpha,\mathbf{0}} \hat{V}^\mathrm{KS},\hat{p}_{\alpha'} \right]
    		\ket{u_{n\mathbf{k}}} \\
    		&\times l_{\mathbf{q}\nu}^2 \frac{M_p}{M_\kappa}e_{\kappa \alpha,\nu}^*(\mathbf{q})e_{\kappa \alpha',\nu}(\mathbf{q})
    		\left[ n_{\mathbf{q}\nu}(T)+\frac{1}{2} \right],
    	\end{aligned}
    \end{align}
    where ${\hat{p}}_\alpha=-i \partial/\partial \hat{r}_\alpha$ is the $\alpha$-th component of the momentum operator.
    \subsection{\label{subsec:2B}The ORDW term and the OSR}
    We then rewrite the DW term by inserting the closure relation $\sum_{m,\mathbf{k}}\ket{\psi_{m\mathbf{k}}}\bra{\psi_{m\mathbf{k}}}=\hat{1}$ into Eq. (\ref{eq:10}), 
    \begin{align} \label{eq:11}
    	\Delta\varepsilon_{n\mathbf{k}}^\mathrm{DW}\left(T\right)=\sum_{m}{\Delta\varepsilon_{n\mathbf{k},m}^\mathrm{ORDW}\left(T\right)},
    \end{align}
	where
	\begin{align} \label{eq:12}
		\begin{aligned}
			\Delta\varepsilon_{n\mathbf{k},m}^\mathrm{ORDW}\left(T\right)=&
			\mathrm{Re} \frac{i}{N_p}
			\sum_{\mathbf{q},\nu}
			\sum_{\kappa,\alpha,\alpha^\prime}
			l_{\mathbf{q}\nu}^2\frac{M_p}{M_\kappa}e_{\kappa\alpha,\nu}^\ast\left(\mathbf{q}\right) \\
			&\times e_{\kappa\alpha^\prime,\nu}\left(\mathbf{q}\right)
			\left[h_{mn,\kappa\alpha}\left(\mathbf{k},\mathbf{0}\right)\right]^\ast \\
			& \times {\left\langle u_{m\mathbf{k}}\middle|{\hat{p}}_{\alpha^\prime}\middle| u_{n\mathbf{k}}\right\rangle}_\mathrm{uc}
			\left[2n_{\mathbf{q}\nu}\left(T\right)+1\right].
		\end{aligned}
	\end{align}	
    Further, by using the relation derived by Lihm and Park~\cite{2020Lihm}
\begin{align} \label{eq:13}
	\sum_{\kappa}{h_{mn,\kappa\alpha}(\mathbf{k},\mathbf{0})}=i\left(\varepsilon_{m\mathbf{k}}-\varepsilon_{n\mathbf{k}}\right){\left\langle u_{m\mathbf{k}}\middle|{\hat{p}}_\alpha\middle| u_{n\mathbf{k}}\right\rangle}_\mathrm{uc},
\end{align}
	Eq. (\ref{eq:12}) is rewritten as
    \begin{align} \label{eq:14}
    	\begin{aligned}
    		\Delta \varepsilon_{n \mathbf{k},m}^{\mathrm{ORDW}}(T)=&
    		-\mathrm{Re} \frac{1}{N_p} \sum_{\mathbf{q},\nu}\sum_{\kappa,\alpha,\kappa^\prime,\alpha^\prime}
    		l_{\mathbf{q}\nu}^2
    		\frac{M_p}{M_\kappa}  e_{\kappa \alpha, \nu}^{*}(\mathbf{q}) \\
    		&\times e_{\kappa \alpha^{\prime}, \nu}(\mathbf{q}) \frac{\left[h_{m n, \kappa \alpha}(\mathbf{k}, \mathbf{0})\right]^{*} h_{m n, \kappa^\prime \alpha^{\prime}}(\mathbf{k}, \mathbf{0})}{\varepsilon_{n\mathbf{k}}-\varepsilon_{m\mathbf{k}}+ i \eta} \\
    		&\times \left[ 2n_{\mathbf{q} \nu}(T)+1 \right].
    	\end{aligned}
    \end{align}
	Here the convergence factor is inserted.
    The momentum operator in Eq. (\ref{eq:12}) simplifies relating the ORDW term with the symmetry of the material and, thus, with the OSR inherent in the system.
    Further, owing to the momentum operator, the ORDW term $\Delta\varepsilon_{n\mathbf{k},m}^\mathrm{ORDW}\left(T\right)$ vanishes for a pair of KS orbitals that are optically forbidden. The optically allowed combination is expressed as \cite{1997Kitaev}
    \begin{align} \label{eq:15}
        \left[\Gamma\left(u_{m\mathbf{k}}\right)\otimes\Gamma\left(u_{n\mathbf{k}}\right)\right]\cap\Gamma\left(x,y,z\right)\neq\emptyset,
    \end{align}
    where $\otimes$ is the operator for the direct product, and $\Gamma\left(u_{n\mathbf{k}}\right)$ and $\Gamma\left(x,y,z\right)$ are the irreducible representations belonging to the $\ket{u_{n\mathbf{k}}}$ and Cartesian coordinates, respectively. The OSR is listed in Table \ref{tab:1} for crystals that have point-group $O_h$ or $T_d$, which include diamond-type and zinc-blende-type materials, respectively.
    \begin{table*}[!hbt]  \centering
        \caption{\label{tab:1} OSR of diamond-type materials (point group $O_h$; left) and zinc-blende-type materials (point group $T_d$; right). $+$ denotes an allowed transition.}
        \begin{tabular}{ccccccccccc} \hline
\hline
            $O_h$ & $A_{1g}$ & $A_{2g}$ & $E_{g}$ & $T_{1g}$ & $T_{2g}$ &
            $A_{1u}$ & $A_{2u}$ & $E_{u}$ & $T_{1u}$ & $T_{2u}$ \\
            \hline
            $A_{1g}$ & & & & & & & & & + & \\
            $A_{2g}$ & & & & & & & & & & + \\
            $E_{g}$ & & & & & & & & & + & + \\
            $T_{1g}$ & & & & & & + & & + & + & + \\
            $T_{2g}$ & & & & & & & + & + & + & + \\
            $A_{1u}$ & & & & + & & & & & & \\
            $A_{2u}$ & & & & & + & & & & & \\
            $E_{u}$ & & & & + & + & & & & & \\
            $T_{1u}$ & + & & + & + & + & & & & & \\
            $T_{2u}$ & & + & + & + & + & & & & & \\
            \hline
\hline
        \end{tabular}
        \begin{tabular}{cccccc} \hline
\hline
            $T_d$ & $A_{1}$ & $A_{2}$ & $E$ & $T_{1}$ & $T_{2}$ \\
            \hline
            $A_1$ & & & & & + \\
            $A_2$ & & & & + & \\
            $E$ & & & & + & + \\
            $T_1$ & & + & + & + & + \\
            $T_2$& + & & + & + & + \\
            \hline
\hline
        \end{tabular}
    \end{table*}
    \subsection{\label{subsec:2C}Computational details}
    We calculated the ORDW terms for several diamond-type and zinc-blende-type materials: C, Si, Ge, SiC, BN, and BP. The calculation was performed based on density functional theory (DFT) \cite{1964Hohenberg,1965Kohn} within the semilocal approximations for the exchange-correlation functional. The generalized gradient approximation (GGA) developed by Perdew, Burke, and Ernzerhof (PBE) \cite{1996Perdew} was used except for Ge, for which the local density approximation (LDA) was used to avoid the bandgap vanishing problem. It should also be noted that the zero-point lattice expansion effect is also important for Ge~\cite{2020Miglio}, but we did not consider it in this paper. We used the optimized norm conserving Vanderbilt (ONCV) pseudopotentials \cite{2013Hamann} developed by Schlipf and Gygi (SG15) \cite{2015Schlipf} and the plane wave cutoff of 60 Ry for wave functions and 240 Ry for charge densities. Furthermore, the electronic and phonon structure was obtained using the Quantum Espresso package~\cite{2009Giannozzi,2017Giannozzi}.

    The primitive cell was optimized theoretically to obtain the lattice constants 6.717, 10.347, 9.522, 8.261, 6.805, and 8.581 $a_0$ for C, Si, Ge, SiC, BN, and BP, respectively, where $a_0$ is the Bohr radius. For the convergence factor $\eta$ of the FM and DW terms, we adopted a value of $0.1~\mathrm{eV}$ following our convergence study~\cite{Supp}. We used the Monkhorst-Pack grids of size $8\times8\times8$ for the electronic and phonon structure calculations, while for the calculation of the FM and DW terms, we used a fine grid of size $20\times 20\times 20$ with Fourier interpolation. We used 15 as the number of bands to sum over in the FM [Eq. (\ref{eq:7})] and DW [Eq. (\ref{eq:11})] terms. See Supplemental Material~\cite{Supp}, which includes Refs.~\cite{2009Giannozzi, 2011Gonze, 2017Giannozzi, 2020Lihm}, for details on the convergence study of the grids, bands, and convergence factor $\eta$.
    \section{\label{sec:3}Results and discussions}
    \subsection{\label{subsec:3A}The DW terms for diamond-type and zinc-blende-type materials}
    We now detail the meaning of the DW term specifically for diamond-type and zinc-blende-type materials. For these materials,
    \begin{align} \label{eq:16}
    	\begin{aligned}
    		& \sum_{\mathbf{q},\nu}l_{\mathbf{q}\nu}^2
    		e_{\kappa \alpha, \nu}^*(\mathbf{q}) e_{\kappa \alpha', \nu}(\mathbf{q})
    		\left[ 2n_{\mathbf{q}\nu}(T)+1 \right] \\
    		&=\delta_{\alpha,\alpha'}
    		\sum_{\mathbf{q},\nu}l_{\mathbf{q}\nu}^2
    		\left| e_{\kappa 1,\nu}(\mathbf{q}) \right|^2
    		\left[ 2n_{\mathbf{q}\nu}(T)+1 \right]
    	\end{aligned}	
	\end{align}
	holds because of the point group symmetry.
    Then, the ORDW term [Eq.(\ref{eq:12})] can be rewritten as
    \begin{align} \label{eq:17}
    	\begin{aligned}
    		\Delta\varepsilon_{n\mathbf{k},m}^\mathrm{ORDW}\left(T\right)= &
    		\mathrm{Re}~i\sum_{\kappa,\alpha} \braket{\Delta \hat{\tau}_\kappa^2}_T
    		\left[h_{mn,\kappa\alpha}\left(\mathbf{k},\mathbf{0}\right)\right]^\ast \\
    		&\times \left\langle u_{m\mathbf{k}}\middle|{\hat{p}}_\alpha\middle| u_{n\mathbf{k}}\right\rangle_\mathrm{uc},
    	\end{aligned}
    \end{align}
	where the MSD of a nucleus $\kappa$ at a temperature $T$ is
	\begin{align} \label{eq:18}
		\braket{\Delta \hat{\tau}^2_\kappa}_T = \frac{M_p}{N_p M_\kappa}\sum_{\mathbf{q},\nu}l_{\mathbf{q}\nu}^2
		\left| e_{\kappa 1,\nu}(\mathbf{q}) \right|^2
		\left[ 2n_{\mathbf{q}\nu}(T)+1 \right].
	\end{align}
    The ORDW term can then be rewritten using Eq. (\ref{eq:13}) as
    \begin{align} \label{eq:19} \begin{aligned}
            \Delta \varepsilon_{n \mathbf{k},m}^{\mathrm{ORDW}}(T) = & (\varepsilon_{m\mathbf{k}}-\varepsilon_{n\mathbf{k}})
            \frac{ \langle \Delta \hat{\tau}_{1}^2 \rangle_T + \langle \Delta \hat{\tau}_{2}^2 \rangle_T}{2} \\
            &\times \sum_{\alpha} \left| \bra{u_{n\mathbf{k}}} \hat{p}_\alpha \ket{u_{m\mathbf{k}}}_\mathrm{uc} \right|^2 \\
            &+\mathrm{Re}~i \frac{ \langle \Delta \hat{\tau}_{1}^2 \rangle_T - \langle \Delta \hat{\tau}_{2}^2 \rangle_T}{2} \\
            &\times \sum_{\alpha}
            \left[ h_{mn,1 \alpha}(\mathbf{k,0})-h_{mn,2 \alpha}(\mathbf{k,0})\right]^* \\
            &\times \bra{u_{m\mathbf{k}}} \hat{p}_\alpha \ket{u_{n\mathbf{k}}}_\mathrm{uc}.
    \end{aligned} \end{align}
    For diamond-type materials, where $\braket{\Delta \hat{\tau}_1^2}_T=\braket{\Delta \hat{\tau}_2^2}_T$ holds, the MSD can be represented by the average of these values denoted as $\braket{\Delta \hat{\tau}^2}_T$. Subsequently, the second term on the right-hand side of Eq. (\ref{eq:19}) vanishes. Therefore, we obtain
    \begin{align} \label{eq:20}
        \Delta \varepsilon_{n \mathbf{k},m}^{\mathrm{ORDW}}(T) \simeq (\varepsilon_{m\mathbf{k}}-\varepsilon_{n\mathbf{k}}) \langle \Delta \hat{\tau}^2 \rangle_T \sum_{\alpha} \left| \bra{u_{n\mathbf{k}}} \hat{p}_\alpha \ket{u_{m\mathbf{k}}}_\mathrm{uc} \right|^2,
    \end{align}
    which is given as the product of the optical transition energy, MSD, and dipole transition probability, and thus the DW term can be evaluated using these known properties of the materials. For zinc-blende-type materials, Eq. (\ref{eq:20}) will be valid when the MSD is nearly symmetric $\braket{\Delta \hat{\tau}_1^2}_T\simeq \braket{\Delta \hat{\tau}_2^2}_T$; the validation is presented in subsequent sections.
    \begin{figure*}[!hbt] \centering
        \includegraphics[width=0.75\textwidth]{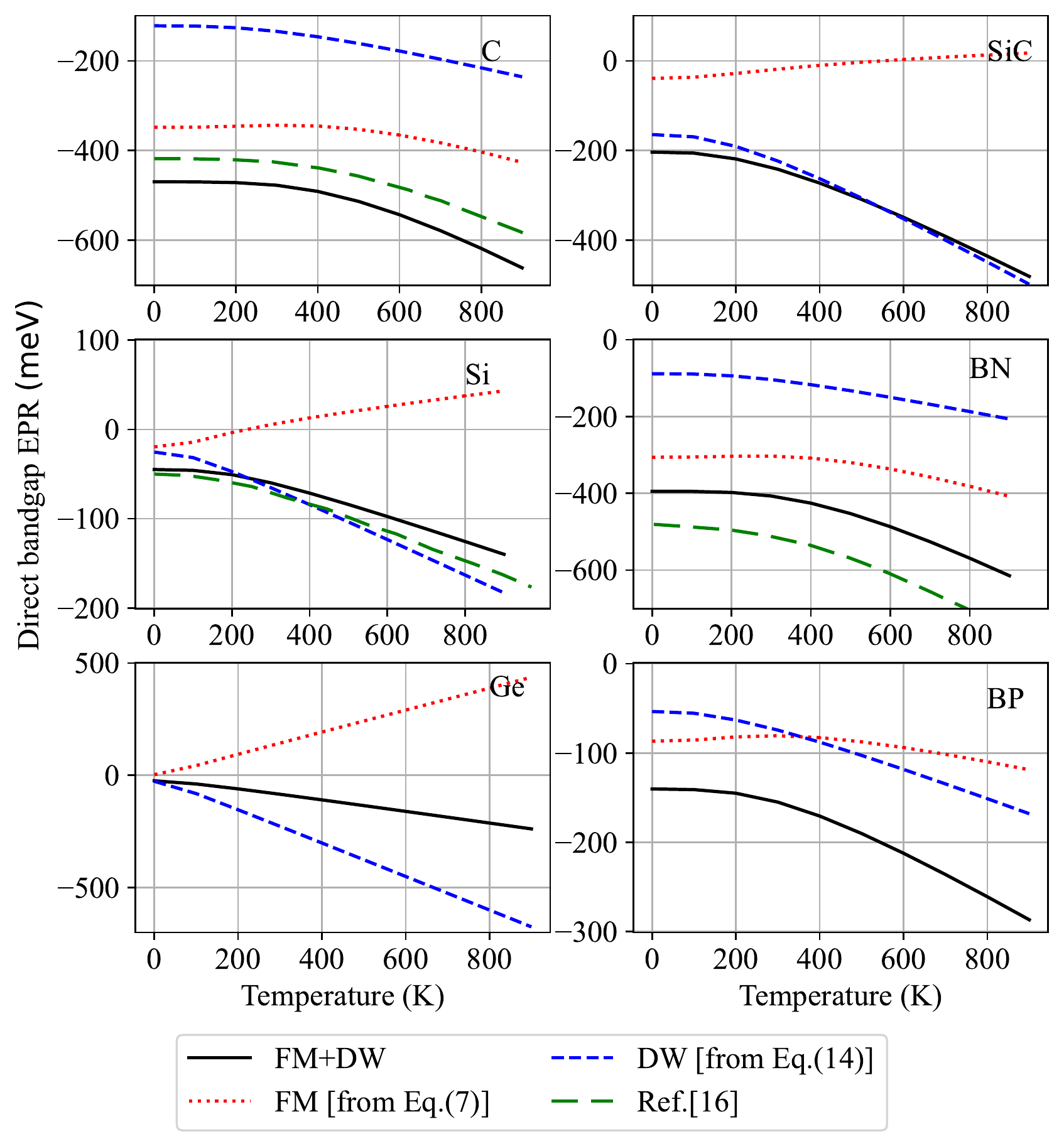}
        \caption{\label{fig:1}(color online): The direct bandgap EPR (black solid line) together with the FM (red dotted line) and DW (blue broken line) terms. The direct bandgap EPR from a previous study \cite{2015Ponce} is also shown (green long-dashed line).}
    \end{figure*}
    \subsection{\label{subsec:3B}Numerical validation of the direct bandgap EPR and ORDW term}
    \begin{table*}[!hbt] \centering
           \caption{\label{tab:2}The direct bandgaps at $\mathbf{k}=\mathbf{0}$ and the corresponding EPR at $T=0~\mathrm{K}$. The first and second columns show our calculated values, the third shows experimental data, and the fourth shows theoretical data with AHC theory.}
        \begin{tabular}{ccccc} \hline \hline
			& Direct bandgap ($\mathrm{eV}$) & EPR (meV)& & \\
            & This work  & This work & Exp. & Theory \\ \hline
            $\mathrm{C}$ & 5.64 & -470 & -320~\cite{1992Logothetidis}, -450~\cite{1992Logothetidis} & -416~\cite{2020Miglio}, -415.8~\cite{2015Ponce} \\
            $\mathrm{Si}$ & 2.55 & -45 & -25~\cite{1987Lautenschlager} & -42~\cite{2020Miglio}, -42.1~\cite{2015Ponce} \\
            $\mathrm{Ge}$ & 2.57 & -25 & -45~\cite{1995Yin} & - \\
            $\mathrm{SiC}$ & 6.14 & -203 & -& -213~\cite{2021Shang} \\
            $\mathrm{BN}$ & 8.86 & -395 & - & -460~\cite{2021Shang}, -502.0~\cite{2015Ponce}\\
            $\mathrm{BP}$ & 3.42 & -140 & - & -101~\cite{2021Shang}\\ \hline
\hline
        \end{tabular}
    \end{table*}
    \begin{table}[!hbt] \centering
        \caption{\label{tab:3}The MSDs at $T=0~\mathrm{K}$ ($10^{-3} a_0^2$). The first and second columns show our calculated values and the third shows the experimental data at $1~\mathrm{K}$. The difference of $1~\mathrm{K}$ is insignificant in this case.}
        \begin{tabular}{cccc} \hline
\hline
            & $\langle \Delta \hat{\tau}_1^2 \rangle_{0\mathrm{K}}$ & $\langle \Delta \hat{\tau}_2^2 \rangle_{0\mathrm{K}}$ &  $\langle \Delta \hat{\tau}^2 \rangle_{1\mathrm{K}}$(Exp.)~\cite{1996Peng} \\ \hline
            $\mathrm{C}$ & 5.723 & 5.723 & 5.816 \\
            $\mathrm{Si}$ & 8.727 & 8.727 & 8.661 \\
            $\mathrm{Ge}$ & 5.319 & 5.319 & 6.065 \\
            $\mathrm{SiC}$ & 7.976 & 5.518 & - \\
            $\mathrm{BN}$ & 6.997 & 5.689 & - \\
            $\mathrm{BP}$ & 9.521 & 5.656 & - \\ \hline
\hline
        \end{tabular}
    \end{table}
    Before examining the applicability of Eq. (\ref{eq:20}), we checked our computational setup. We have computed the direct bandgap EPR at $\mathbf{k}=\mathbf{0}$ in the temperature range 0--900~$\mathrm{K}$ and the MSD at $0~\mathrm{K}$. Figure \ref{fig:1} shows that our EPR value is reasonably consistent with the prior calculations done for C, Si, and BN \cite{2015Ponce}. The minor difference may have originated from different computational conditions. Table \ref{tab:2} shows that, the direct bandgap EPR at 0 K is consistent with experimental and theoretical values, although typically differing by 20--70~$\mathrm{meV}$. Table \ref{tab:3} shows that the calculated MSD at $0~\mathrm{K}$ differs from the experimental value by only $0.1 \times{10}^{-3}a_0^2$ for C and Si, while the difference is much larger, $0.7 \times{10}^{-3}a_0^2$, for Ge. The larger difference found for Ge may be due at least partially to the LDA used for the exchange-correlation potential. As mentioned above, we have used LDA to avoid too small a bandgap given by GGA, and we intend to carry out more accurate calculations in a future study. Despite a certain amount of inaccuracy in the present calculation, it is worth emphasizing that the magnitude of EPR is one order of magnitude larger for C compared with Si and Ge, and explaining this material dependence is the main objective of our study.
    \begin{figure}[!hbt]  \centering
        \includegraphics[width=0.3\textwidth]{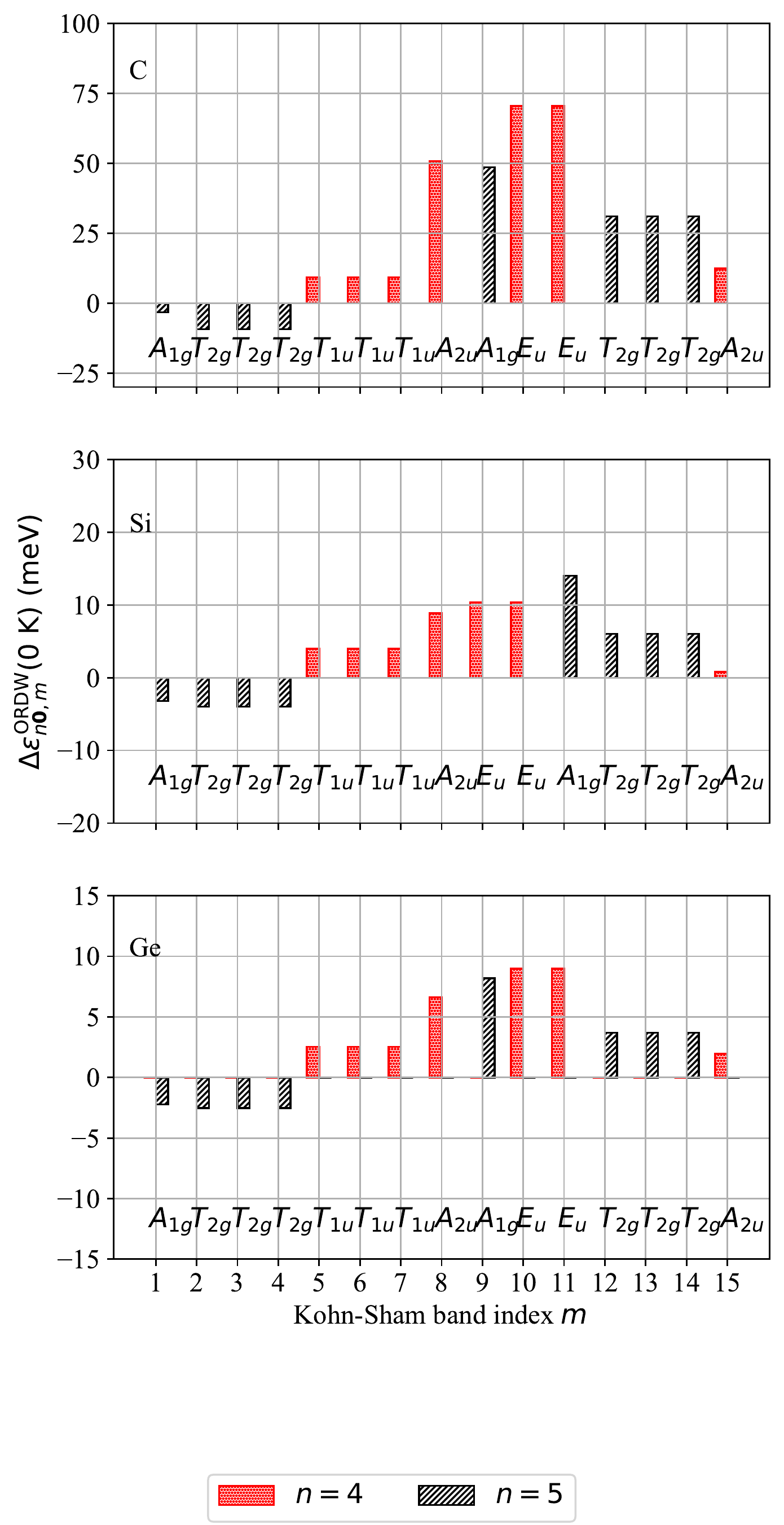}
        \caption{\label{fig:2}(color online) The ORDW terms, $\Delta \varepsilon_{n\mathbf{0},m}^\mathrm{ORDW}(0~\mathrm{K})$, for C, Si, and Ge. Plots are given for the valence top, $n=4$ (shaded with red), and the conduction bottom, $n=5$ (hatched with black). The horizontal axis is the KS band index $m$ whose irreducible representation is also shown. Note that the successive states with the same irreducible representation, $E$ and $T$, are doubly and triply degenerate, respectively, and the value of $\Delta \varepsilon_{n\mathbf{0},m}^\mathrm{ORDW}(0~\mathrm{K})$ has been averaged over the degenerate states. }
    \end{figure}
    As an additional validation of the numerical accuracy, we examined the OSR. Figure \ref{fig:2} shows the plot of the ORDW terms $\Delta \varepsilon_{n\mathbf{0},m}^\mathrm{ORDW}(0~\mathrm{K})$ of diamond-type materials. As a demonstration, we show the results only for $n=4$ (the valence top) and $n=5$ (the conduction bottom) with $m=$1--15 being considered. We find that the value is zero for all the optically forbidden pairs of $n$ and $m$ (see Table \ref{tab:1} left for the OSR). 
    
\begin{figure}[!hbt] \centering
        \includegraphics[width=0.3\textwidth]{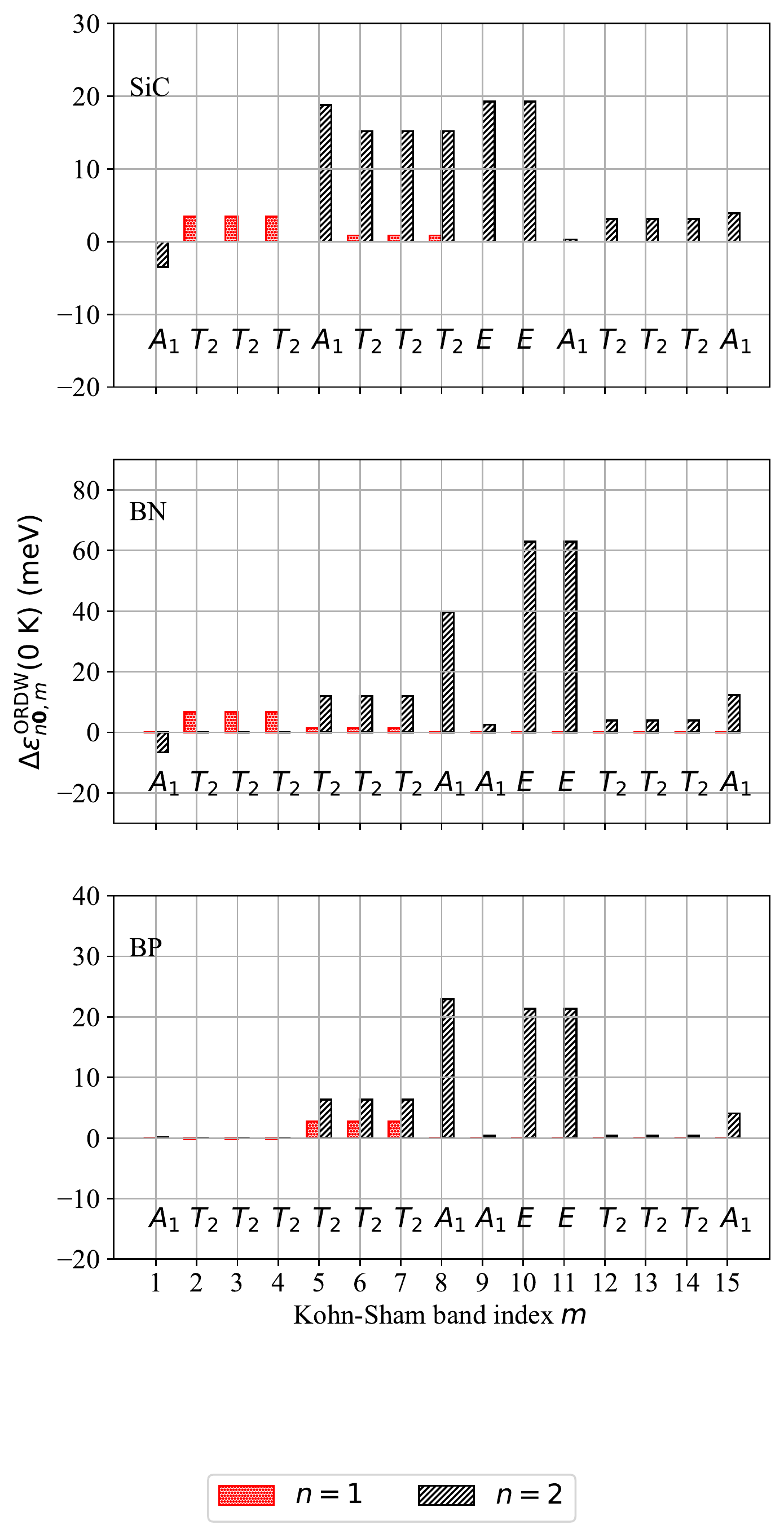}
        \caption{\label{fig:3} (color online) The ORDW terms, $\Delta \varepsilon_{n\mathbf{0},m}^\mathrm{ORDW}(0~\mathrm{K})$, for SiC, BN, and BP. Plots are shown for the lowest valence, $n=1$ (shaded with red) and the valence top, $n=2$ (hatched with black). Details of the notation are the same as in Fig. \ref{fig:2}.}
    \end{figure}

    A similar validation was performed for zinc-blende-type materials, the results of which are shown in Fig. \ref{fig:3} only for $n=1$ (the lowest valence) and $n=2$ (the valence top). The results indicate that the value of $\Delta \varepsilon_{n\mathbf{0},m}^\mathrm{ORDW}(0~\mathrm{K})$ is zero for all the optically forbidden pairs without exception (see Table \ref{tab:1} right for the OSR).
    \subsection{\label{subsec:3C}The origin of the material dependence of the ORDW term}
\begin{figure*}[!hbt] \centering
        \includegraphics[width=0.75\textwidth]{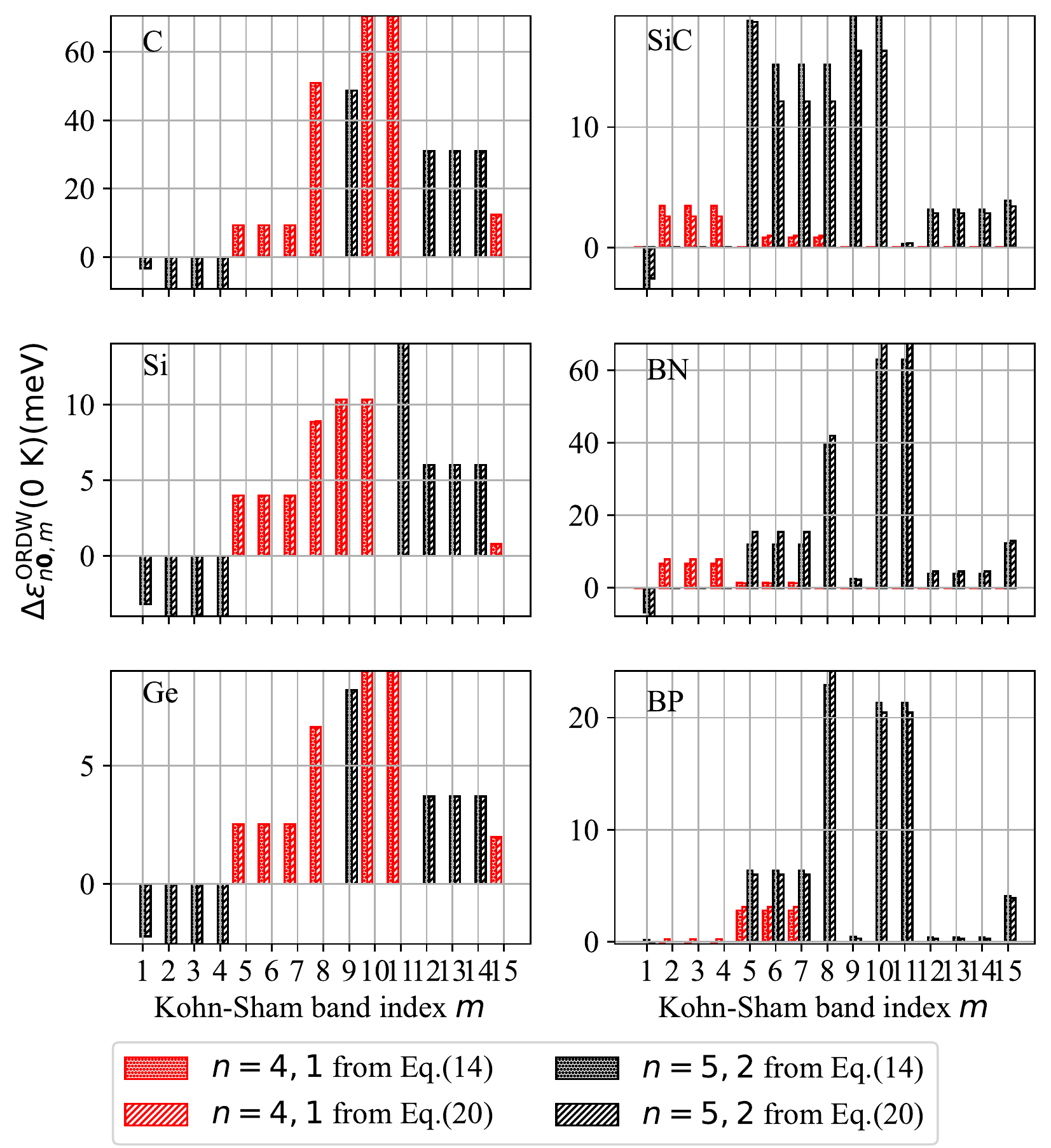}
        \caption{\label{fig:4}(color online) Comparison of the ORDW terms $\Delta \varepsilon_{n\mathbf{0},m}^\mathrm{ORDW}(0~\mathrm{K})$ obtained with the original formula [Eq. (\ref{eq:14})] with those obtained with the approximate one [Eq. (\ref{eq:20})]. Details of the notation are similar to those in Figs. \ref{fig:3} and \ref{fig:4}.}
    \end{figure*}

    Now we compare the ORDW term obtained using the original formula [Eq. (\ref{eq:14})] with that obtained using the approximate one [Eq. (\ref{eq:20})], as plotted in Fig. \ref{fig:4}. In the diamond-type case, perfect agreement can be found between the rigorous and the approximate formulae. In the zinc-blende case, the match is not perfect, but the difference is within several percent.

    It is noteworthy that the ORDW terms are larger for C than for Si and Ge by an order of magnitude. This striking material dependence primarily originates from the generally larger spacing between the energy bands $\varepsilon_{m\mathbf{k}}-\varepsilon_{n\mathbf{k}}$ and the momentum matrix element $ |\bra{u_{m\mathbf{k}}}\hat{p}_\alpha \ket{u_{n\mathbf{k}}}_\mathrm{uc}|$ for C and not from its MSD $\braket{\Delta \hat{\tau}^2}_T$, which is different from that of Si and Ge by up to about 50 percent (Table \ref{tab:3}).
    Table \ref{tab:4} shows that the energy and momentum of C are greater than those of Si and Ge for almost all transitions.
        \begin{table}[!hbt] \centering
    	\caption{\label{tab:4}Momentum matrix elements 
    		$\sum_\alpha |\bra{u_{m\mathbf{0}}}\hat{p}_\alpha \ket{u_{n\mathbf{0}}}_\mathrm{uc}|^2~(n=4,5)$ of diamond-type materials ($a_0^{-2}$). Values are averaged over degenerate states. Transitions are from the valence top $(T_{2g})$ or conduction bottom $(T_{1u})$ states to the nearest singly ($A$), doubly ($E$), or triply ($T$) degenerate states. Values in parentheses are the transition energies (eV).}
    	\begin{tabular}{cccc} \hline
    		\hline
    		& $T_{2g} \to T_{1u}$  & $T_{2g} \to A_{2u}$ & $T_{2g} \to E_u$ \\ \hline
    		$\mathrm{C}$ & 0.096~(5.64) & 0.655~(13.6) & 0.230~(26.8) \\
    		$\mathrm{Si}$ & 0.060~(2.55) & 0.333~(3.06) & 0.077~(7.66) \\
    		$\mathrm{Ge}$ & 0.062~(2.57) & 0.375~(3.32) & 0.080~(10.5) \\ \hline
    		& $T_{1u} \to A_{1g} $  & $T_{1u} \to T_{2g} $ & $T_{1u} \to A_{1g}$ \\ \hline
    		$\mathrm{C}$ & 0.021~(-27.2) & 0.096~(-5.64) & 0.615~(13.8) \\
    		$\mathrm{Si}$ & 0.026~(-14.3) & 0.060~(-2.55) & 0.304~(5.30) \\
    		$\mathrm{Ge}$ & 0.022~(-18.6) & 0.062~(-2.57) & 0.341~(4.52) \\ \hline
    		\hline
    	\end{tabular}
    \end{table}
    
    Although this trend is applicable to diamond-type materials where the material dependence is large, it is not necessarily true when discussing zinc-blende-type materials with minor material dependence (Fig. \ref{fig:4}). In addition, the MSD is considerably asymmetric in $\kappa$ for SiC and BP (Table \ref{tab:3}), contrary to the assumption made in deriving Eq. (\ref{eq:20}). However, generally large values for the ORDW term are found for BN compared with those of SiC and BP, and thus this material dependence is correlated to the larger spacing of the energy bands and momentum matrix elements of BN.
    \begin{figure*}[!hbt] \centering
        \includegraphics[width=0.75\textwidth]{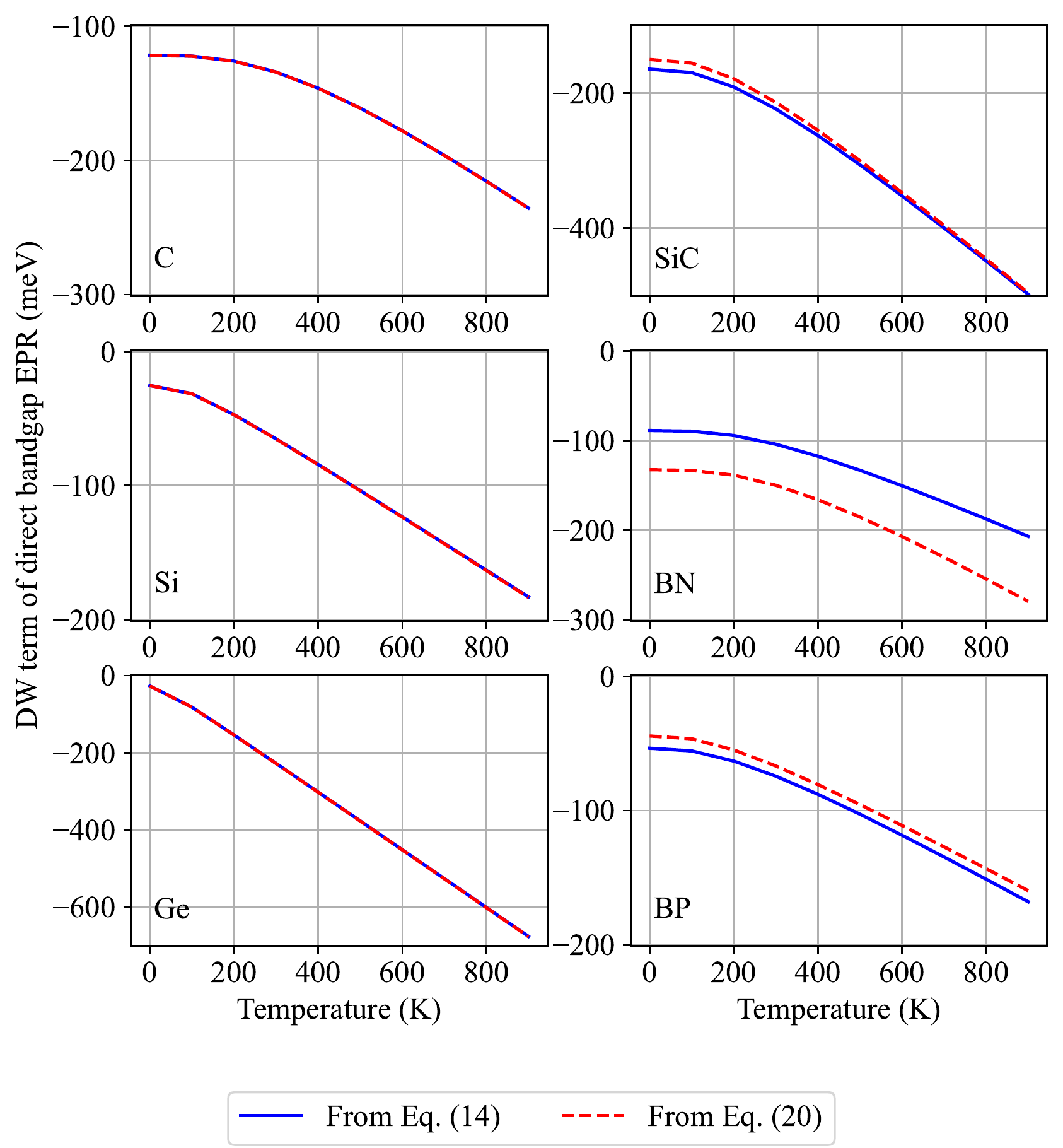}
        \caption{\label{fig:5}(color online): The DW terms of the direct bandgap EPR $\Delta \varepsilon_{5\mathbf{0}}^\mathrm{DW}(T)- \Delta \varepsilon_{4\mathbf{0}}^\mathrm{DW}(T)$ obtained by summing the first $15$ bands, or $m=1$ to $15$, using the original formula [Eq. (\ref{eq:14})] and those obtained using the approximate formula [Eq. (\ref{eq:20})]. The original is plotted with blue solid line while the approximate is plotted with red broken line.}
    \end{figure*}
    Hence, the manner in which the approximate ORDW terms may be summed up to recover the DW term is of interest. However, strict investigation is hampered by the slow convergence of Eq. (\ref{eq:11}). We therefore compared the value obtained by summing the first 15 orbitals using the original formula [Eq.(\ref{eq:14})] with that obtained using the approximate formula [Eq.(\ref{eq:20})]. The DW terms of the direct bandgap EPR obtained from these formulae are plotted in Fig. \ref{fig:5}. The difference between the two is zero for diamond-type materials but not for zinc-blende-type materials. The difference in behavior between diamond-type and zinc-blende-type materials is due to the approximation $\braket{\Delta \hat{\tau}_1^2}_T=\braket{\Delta \hat{\tau}_2^2}_T$.
    Finally, we discuss finite temperature. Because Eq. (\ref{eq:20}) can be rewritten as $\Delta \varepsilon_{n\mathbf{k}, m}^\mathrm{ORDW}(T) \simeq \braket{\Delta \hat{\tau}^2}_T / \braket{\Delta \hat{\tau}^2}_{0\mathrm{K}} \Delta \varepsilon_{n\mathbf{k}, m}^\mathrm{ORDW}(0~\mathrm{K})$, the temperature dependence of the DW term is highly dependent on the temperature dependence of the MSD.
    However, the material dependence of $\Delta \varepsilon_{n\mathbf{k}}^\mathrm{DW}(0~\mathrm{K})$ affects it as a multiplier.
    \section{\label{sec:4}Conclusion}
    We reformulated the DW term of the AHC theory to decompose it into the ORDW terms using the formula recently derived by Lihm and Park~\cite{2020Lihm}, thereby relating the DW term with the optical dipole transition. For diamond- and zinc-blende-type materials, we found that the ORDW term can be factorized into the difference in the KS eigenvalues, the MSD, and the dipole transition probability, as shown in Eq. (\ref{eq:20}) under the approximation of symmetric MSD for the zinc-blende-type. Further, the OSR and approximate factorization were numerically examined for C, Si, Ge, SiC, BN, and BP using first-principles DFT calculations and were found to reproduce the non-approximated results perfectly for diamond-type materials and reasonably well for zinc-blende-type materials. It is therefore possible to estimate the magnitude of the DW term from the known properties of materials before performing elaborate EPI calculations.
    In addition, it is possible to ascribe the larger DW term of C to the larger separation of the electronic bands and momentum matrix elements using our formula.
    In this way, the DW term is given a physical meaning and its material dependence is given an intuitive explanation. Although the applicability of our results is limited to crystal structures with high symmetry, our results can be generalized to estimate the DW term for materials with low anisotropy. Furthermore, although we have focused only on the DW term, further examination of the whole theory may be important to deepen our understanding of the bandgap EPR.
    
    \begin{acknowledgments}
    	The calculations were performed with the facilities of the Supercomputer Center, the Institute for Solid State Physics, the University of Tokyo.
    \end{acknowledgments}
    \bibliography{main}
\end{document}


\makeatletter
	\renewcommand{\@cite}[1]{[{\color{blue}S#1}]}
	\renewcommand{\@biblabel}[1]{[S#1]}
	\makeatother
	\title{Supplemental Material: Optical representation of thermal nuclear fluctuation effect on band-gap renormalization}
	\author{Kohei Ishii, Jun Haruyama, and Osamu Sugino}
	\affiliation{Institute for Solid State Physics, The University of Tokyo, Kashiwa, Chiba 277-8581, Japan}
	\maketitle
	\section{Additional computation information}
	The Fan-Migdal (FM) and Debye-Waller (DW) terms in the main text [Eqs. (7, 11)] are written by the summation of band indexes $m$, so the band cutoff $M_c$ must be determined. In the main text, we set $M_c = 15$ and ignored the contribution from $m>M_c$. However, in Quantum Espresso (QE)~\cite{2009Giannozzi,2017Giannozzi}, the contribution from $m>M_c$ is also calculated using the Sternheimer method extended for AHC theory~\cite{2011Gonze}.
	Here, we call the former the lower FM (lFM) term
	$\Delta E_g^\mathrm{lFM}(T) = \Delta \varepsilon^\mathrm{lFM}_{5\mathbf{0}}(T)- \Delta \varepsilon^\mathrm{lFM}_{4\mathbf{0}}(T)$ and the latter the upper FM (uFM) term
	$\Delta E_g^\mathrm{uFM}(T) = \Delta \varepsilon^\mathrm{uFM}_{5\mathbf{0}}(T)- \Delta \varepsilon^\mathrm{uFM}_{4\mathbf{0}}(T)$.
	The calculation cost of the uFM term is more expensive than that of the lFM term. 
	Also, the DW term $\Delta E_g^\mathrm{DW}(T) = \Delta \varepsilon^\mathrm{DW}_{5\mathbf{0}}(T)- \Delta \varepsilon^\mathrm{DW}_{4\mathbf{0}}(T)$ is represented by Eq. (10)~\cite{2020Lihm} in QE and does not include band summation.
	Its calculation cost is similar to that of uFM.
	
	We implemented additional code based on QE for calculating original and approximate ORDW term [Eqs. (14, 20)].
	We define them as, 
	$\Delta E_g^\mathrm{ORDW}(T)= \sum_{m=1}^{M_c} \left[
	\Delta \varepsilon^\mathrm{ORDW}_{5\mathbf{0},m}(T) - \Delta \varepsilon^\mathrm{ORDW}_{4\mathbf{0},m}(T)
	\right]$ and $\Delta E_g^\mathrm{ORDW(app.)}(T)= \sum_{m=1}^{M_c} \left[
	\Delta \varepsilon^\mathrm{ORDW(app.)}_{5\mathbf{0},m}(T) - \Delta \varepsilon^\mathrm{ORDW(app.)}_{4\mathbf{0},m}(T)
	\right]$.
	Then, the direct bandgap EPR can be calculated in three ways,
	\begin{align}
		\tag{S1}
		\Delta E_g(T)& =
		\Delta E_g^\mathrm{lFM}(T) + \Delta E_g^\mathrm{uFM}(T)+ \Delta E_g^\mathrm{DW}(T)  \\
		\tag{S2}
		\Delta E_g^\mathrm{OR}(T) & =
		\Delta E_g^\mathrm{lFM}(T)  +  \Delta E_g^\mathrm{ORDW}(T) \\
		\tag{S3}
		\Delta E_g^\mathrm{OR(app.)}(T) &  = \Delta E_g^\mathrm{lFM}(T)  +  \Delta E_g^\mathrm{ORDW(app.)}(T).
	\end{align}
	$\Delta E_g(T)$ here is not calculated in the main text, but it is the most accurate method and can be used as a reference value to check the accuracy of our data.
	\section{Convergence study on $\mathbf{q}$ and $\eta$}
\begin{figure}[H] \centering
		\includegraphics[width=\textwidth]{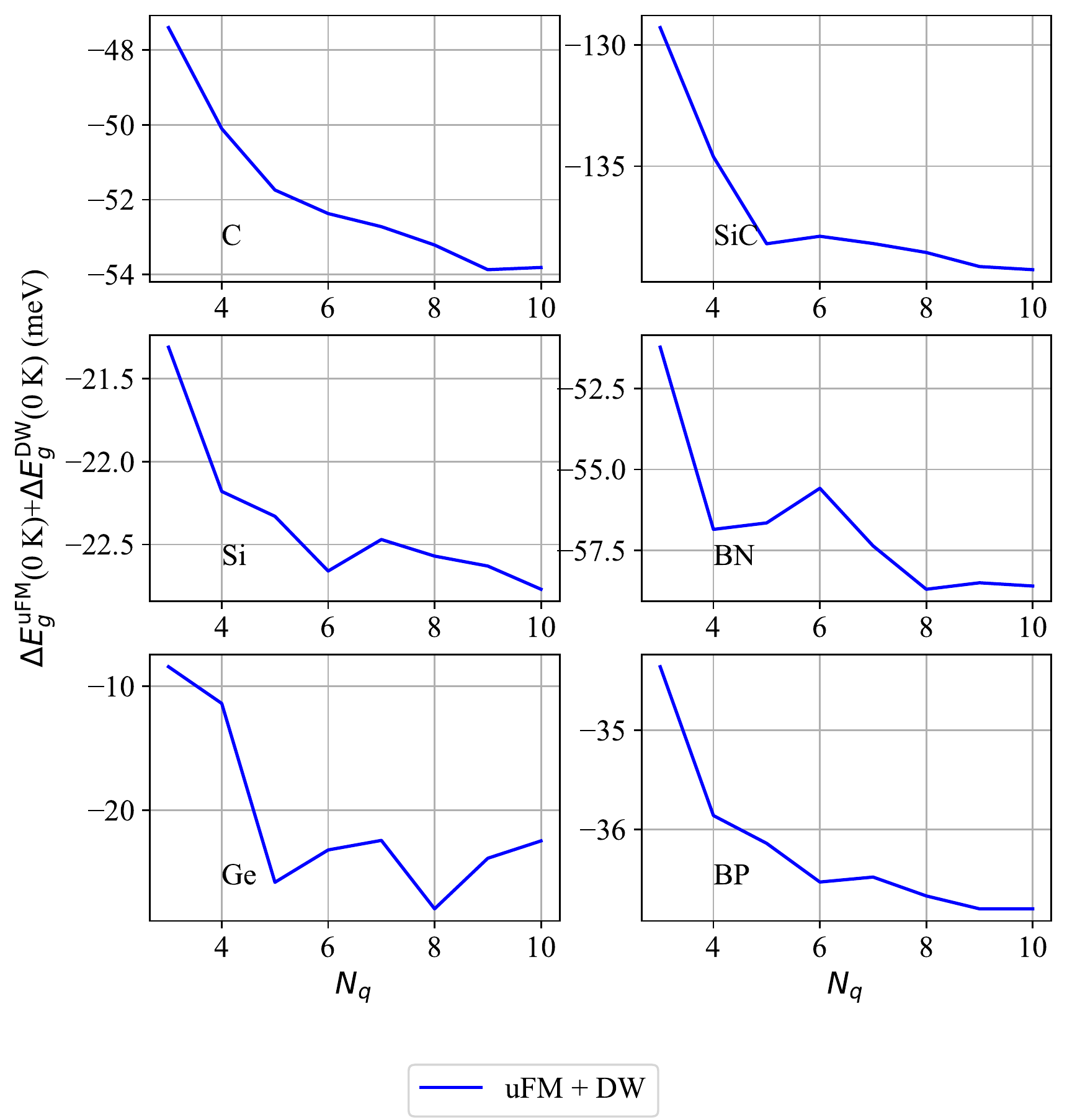}
		\caption{\label{fig:1}$N_q$ dependence of the sum of the uFM and DW terms at $T=0~\mathrm{K}$.}
	\end{figure}

\begin{figure}[H] \centering
		\includegraphics[width=\textwidth]{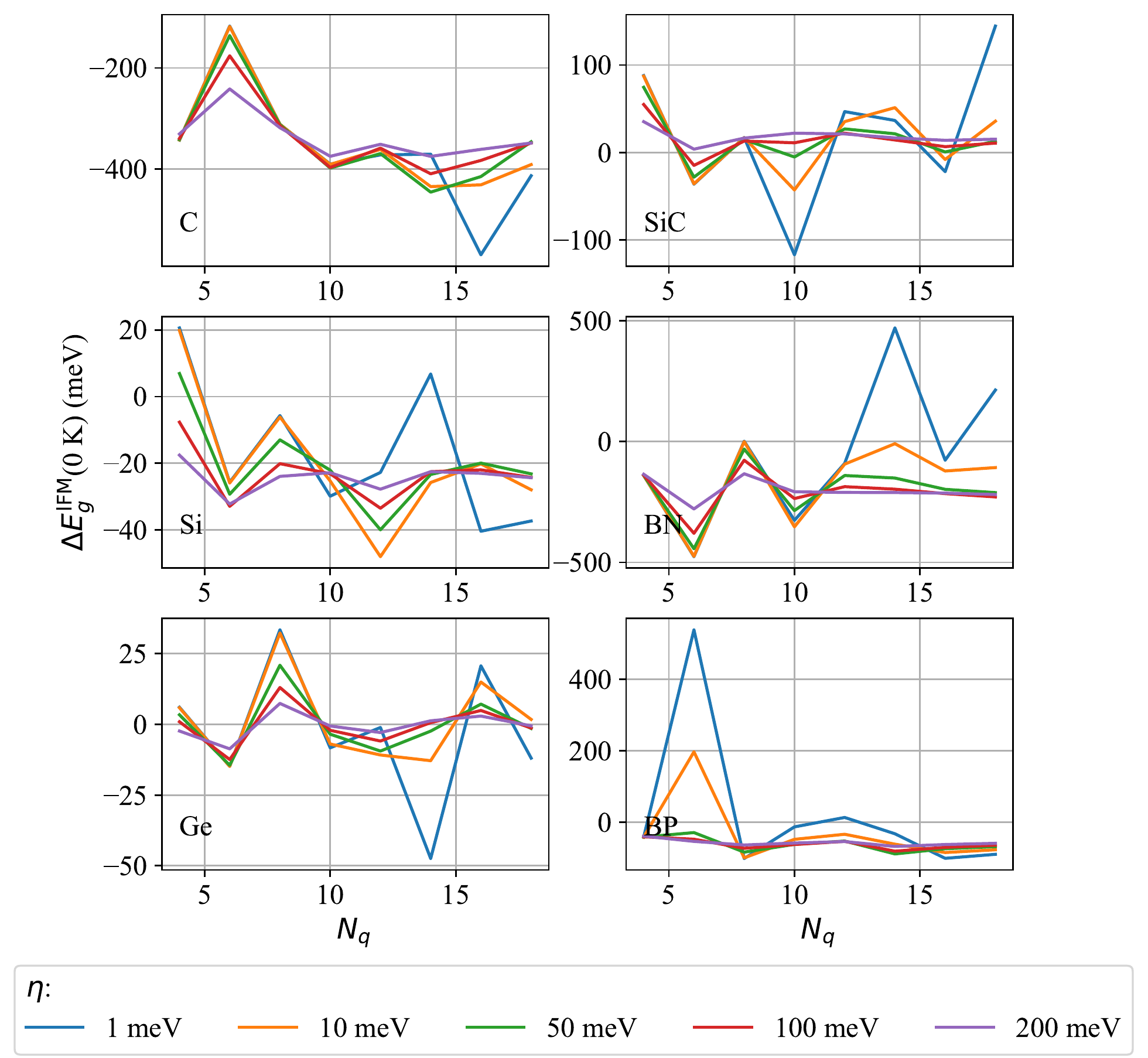}
		\caption{\label{fig:2} 
			$N_q$ dependence of the lFM term at $T=0~\mathrm{K}$ for various $\eta$. The fixed band number is $M_c = 30$.}
	\end{figure}

\begin{figure}[H] \centering
		\includegraphics[width=\textwidth]{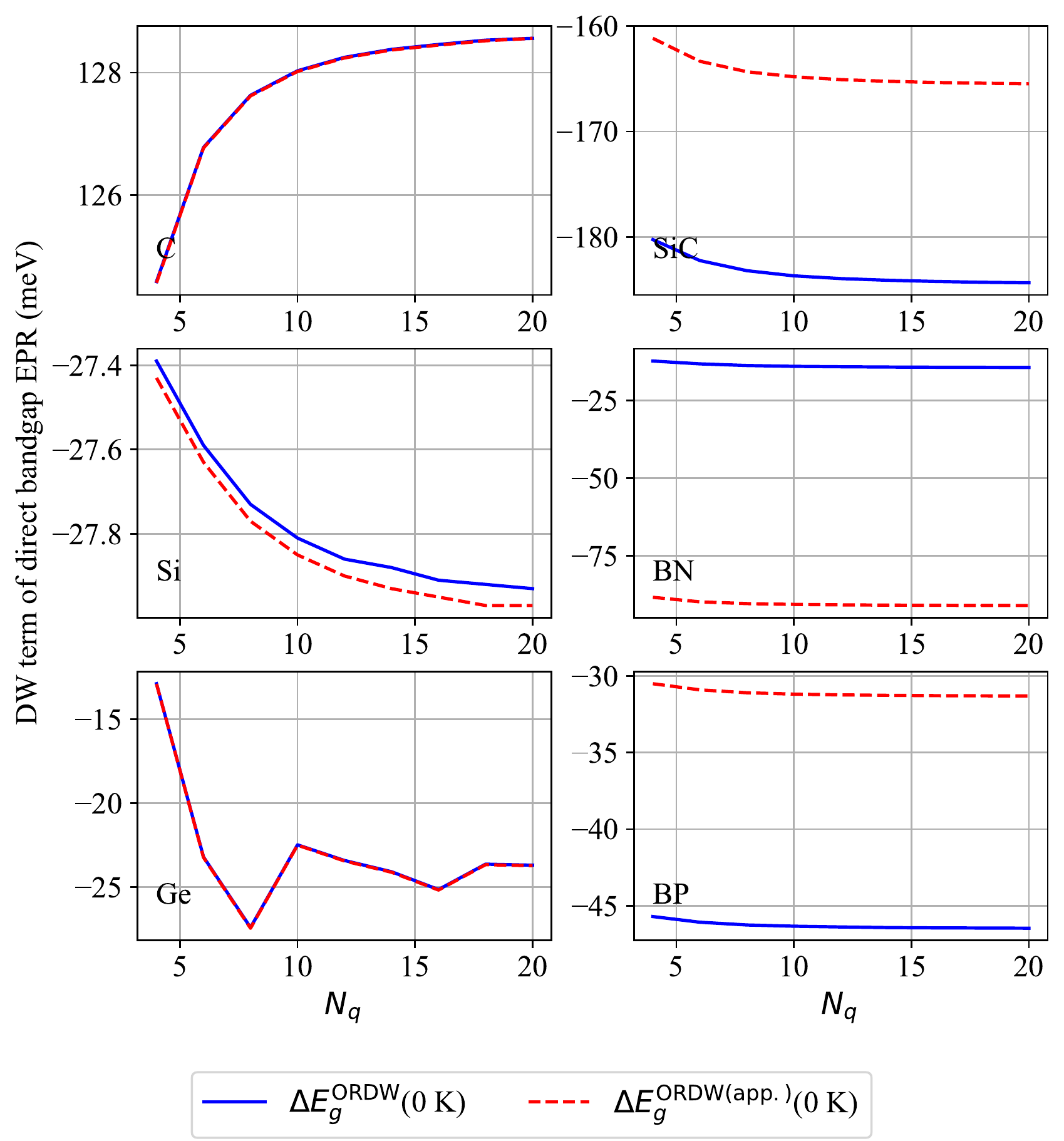}
		\caption{\label{fig:3} 
			$N_q$ dependence of the DW term at $T=0~\mathrm{K}$. The fixed band number is $M_c = 30$.}
	\end{figure}

	We define the size of $\mathbf{q}$ grid as $N_q \times N_q \times N_q$.
	First, we checked the $N_q$ dependence of the uFM and DW terms at $T=0~\mathrm{K}$.
	We plotted $\Delta E_g^\mathrm{uFM}(T) + \Delta E_g^\mathrm{DW}(T)$ [Fig. \ref{fig:1}].
	The value is converged at $N_q=8$.
	Next, we checked the $N_q$ dependence of $\Delta E_g^\mathrm{lFM}(T)$, $\Delta E_g^\mathrm{ORDW}(T)$, and $\Delta E_g^\mathrm{ORDW(app.)}(T)$~[Figs. \ref{fig:2}, \ref{fig:3}].
	The value in the main text, $\eta=0.1~\mathrm{eV}$, is converged at a dense $\mathbf{q}$ grid.
	The lFM and DW terms are converged at $N_q=20$.
	\section{Convergence study of $M_c$}
\begin{figure}[H] \centering
		\includegraphics[width=\textwidth]{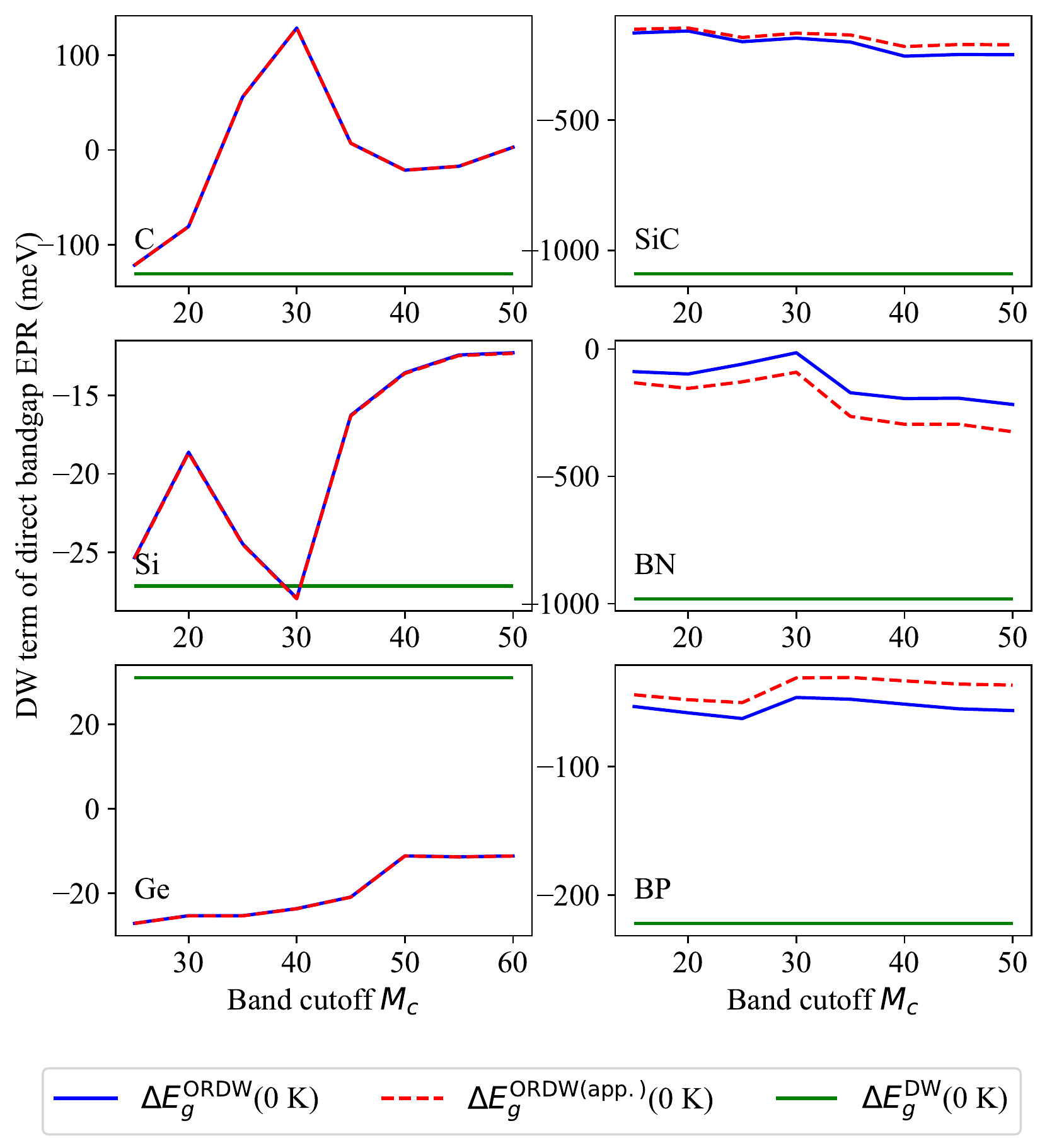}
		\caption{\label{fig:4}
			$M_c$ dependence of the DW term at $T=0~\mathrm{K}$.
			The coarse and fine grid are $8\times 8 \times 8$ and $20\times 20 \times 20$, respectively.}
	\end{figure}

	First, we compared  $\Delta E_g^\mathrm{ORDW}(T)$ and $\Delta E_g^\mathrm{ORDW(app.)}(T)$ to $\Delta E_g^\mathrm{DW}(T)$ [Fig. \ref{fig:4}].
	Although convergence with respect to $M_c$ is difficult, the approximation we claim in the paper is robust with respect to the number of bands for diamond-type materials.
	
\begin{figure}[H] \centering
		\includegraphics[width=\textwidth]{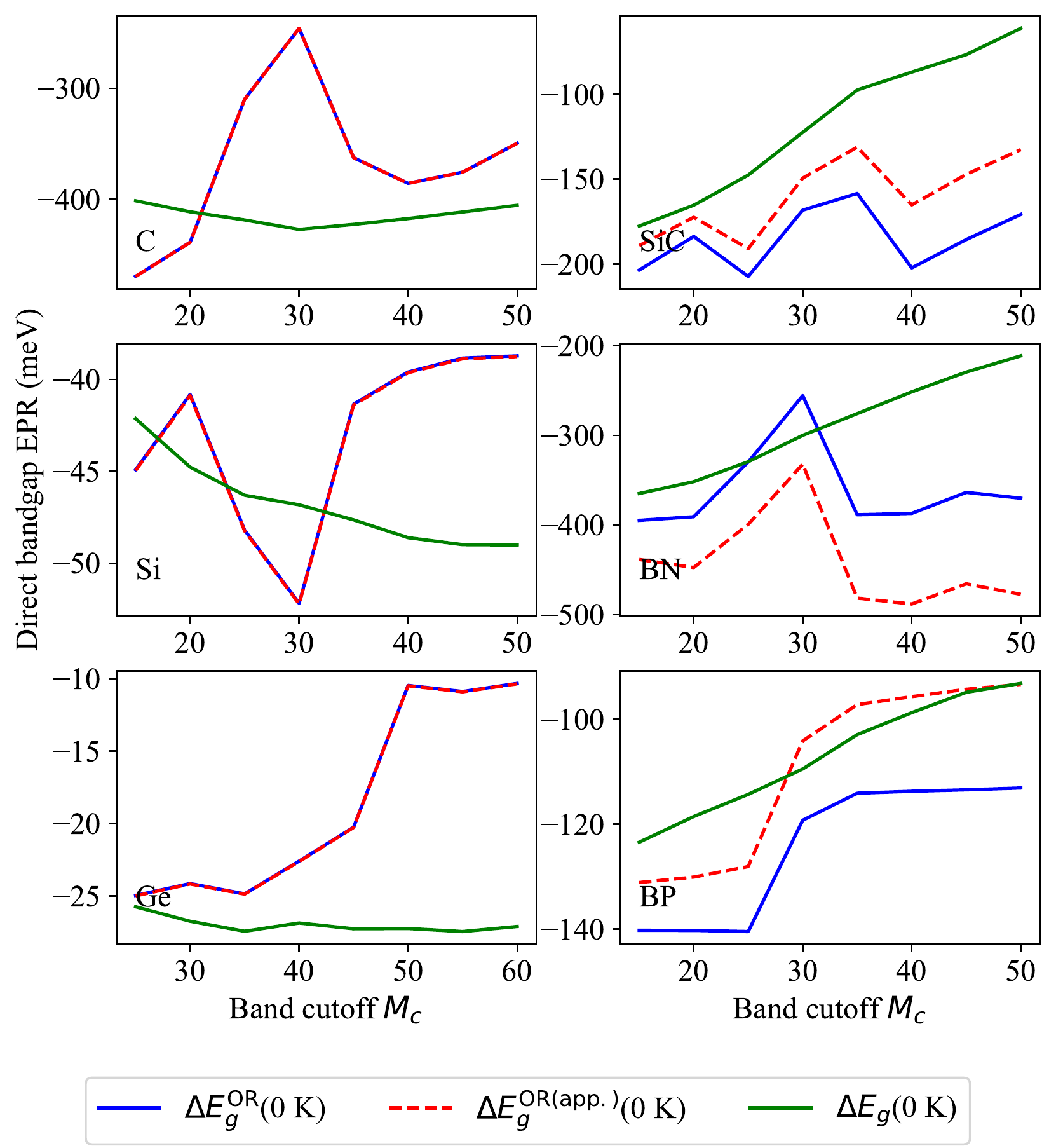}
		\caption{\label{fig:5}
			$M_c$ dependence of EPR at $T=0~\mathrm{K}$.
			The coarse and fine grids are $\mathbf{q}:8\times 8 \times 8$ and $\mathbf{q}:20\times 20 \times 20$, respectively. $\eta=100~\mathrm{meV}$.}
	\end{figure}

	Additionally, we compared $\Delta E_g^\mathrm{OR}(T)$ and $\Delta E_g^\mathrm{OR(app.)}(T)$ 
	to $\Delta E_g(T)$~[Fig. \ref{fig:5}].
	The difference between $\Delta E_g^\mathrm{OR}(T)$ and $\Delta E_g^\mathrm{OR(app.)}(T)$ is almost zero for all $M_c$ for C, Si, and Ge, but is finite and increasing for SiC, BN, and BP.
	Further, the difference between $\Delta E_g^\mathrm{OR}(T)$ or $\Delta E_g^\mathrm{OR(app.)}(T)$ and $\Delta E_g(T)$ generally becomes worse at large $M_c$.
	Therefore, our data in the main text ($M_c=15$) are reasonable, although they have non-negligible band truncation error. 
	\bibliography{supp}